# Simultaneous inversion of layered velocity and density profiles using Direct Waveform Inversion (DWI): 1D case

Zhonghan Liu[1], Yingcai Zheng[1], Hua-Wei Zhou[1]

[1] Department of Earth and Atmospheric Sciences, the University of Houston

# ABSTRACT

We propose and test the Direct Waveform Inversion (DWI) scheme to simultaneously invert for layered velocity and density profiles, using reflection seismic waveforms recorded on the surface. The recorded data include primary reflections and interbed multiples. DWI is implemented in the time-space domain. By explicitly enforcing the wavefield time-space causality, DWI can recursively determine the subsurface seismic structure in a local fashion for both sharp interfaces as well as the properties of the layers, from shallow to deep depths. DWI is different from the layer stripping methods in the frequency domain. By not requiring a global initial model, DWI avoids many nonlinear optimization problems, such as the local minima and the need for an accurate initial model in most full waveform inversion schemes. Using two numerical tests here, we demonstrate that the DWI scheme for multi-parameter seismic inversion for velocity and density of layered media, using plane waves of different incident angles. This represents a significant step comparing with the previous DWI only inverting for the velocity.

## Introduction

Seismic full waveform inversion (FWI), formulated originally by (Lailly, 1983; Tarantola, 1984), is a powerful process in subsurface velocity model building. The goal of FWI is to find a model such that the model-predicted waveforms fit the observed waveforms. Since FWI is an iterative gradient-based method, its success depends on how much the initial model differs from the true model (Virieux and Operto, 2009)**.** The limitation of the iterative FWI scheme was recognized early on (Gauthier *et al.*, 1986; Tarantola, 1986; Mora, 1987; Bourgeois *et al.*, 1989). Tarantola (2005, p.128) pointed out that the local Fréchet gradient used in FWI was equivalent to the single scattering Born approximation. Therefore the performance of FWI relies on an accurate and long-wavelength initial velocity model in which case the Born approximation is more accurate. However, in the Born single scattering, there is a linear correspondence between low-frequency seismic data and low-wavenumber/large-scale structures (Wu and Zheng, 2014). However, due to the lack of low-frequency content (< 5 Hz) in most reflection seismic data, most developments in FWI have been focusing on how to recover large-scale structural information without using low-frequency data. These developments include, for example, the Laplace FWI (Shin and Cha, 2008; Shin and Ha, 2008; Kim *et al.*, 2013), envelope inversion (Wu *et al.*, 2014; Luo and Wu, 2015; Chen *et al.*, 2018), intensity inversion (Liu *et al.*, 2018; Liu *et al.*, 2020), and the FWI using deep learning techniques (Richardson, 2018).

To circumvent the challenges in FWI, we proposed an alternative waveform inversion scheme (Liu and Zheng, 2015; Liu and Zheng, 2017), called the direct waveform inversion (DWI), to invert for subsurface models without the need for a global initial model. DWI combines seismic imaging and velocity model building into one single process. In order to use DWI, it is necessary for the input seismic data to include both free-surface and inter-bed multiples. Using surface recorded reflection seismic data, DWI is able to deliver accurate P-wave velocity inversion results without using a global initial model, for both 1D and 2D layered scalar (i.e., no density variation) models (Zheng and Liu, 2020). Without a global model, DWI inverts the model from shallow to deep depths. In this regard, DWI is similar to the layer-stripping methods (Claerbout, 1976) and the approach by Goupillaud (1961). However, there are important differences in the methods, in particular the explicit use of the time-space causality in DWI.

The purpose of this paper is to examine the DWI's potential to simultaneously invert for velocity and density profiles for a 1d layered medium. It is noted that DWI is not constrained to 1d layered cases. Zheng and Liu (2020) demonstrated DWI also worked for 2d models for the velocity inversion. Multi-parameter attributes of rock strata are important for understanding subsurface properties and reservoir characterization. We have seen that multi-parameter FWI methods had also been proposed to invert for not only the P-wave velocity, but also the S-wave velocity, density, and the seismic anisotropy parameters (Sears *et al.*, 2008; Brossier *et al.*, 2009; Jeong *et al.*, 2012; Warner *et al.*, 2013; Alkhalifah and Plessix, 2014). In the following, we will review the basic idea of DWI in velocity inversion (Liu and Zheng, 2015; 2017), then extend the DWI formulation to the simultaneous inversion for layered velocity and density structures. Two numerical examples are given to demonstrate our methodology.

## 1D DWI with constant density throughout the model

In this section, we briefly summarize the DWI procedure for inverting the sound wave velocity in a layered and horizontally stratified medium that has a uniform density throughout the model (Figure 1). In the next section, we will modify it to include density inversion.

DWI explicitly uses the time-space causality property of the wavefield in the inversion. Starting from the source-receiver layer near the surface, we recursively (not iteratively) build the model downward by fitting the earliest parts of the waveforms of pairs of source-receivers of short (or zero) offsets. We then extrapolate the sources and receivers downward to the bottom of the inverted region, and repeat the process.

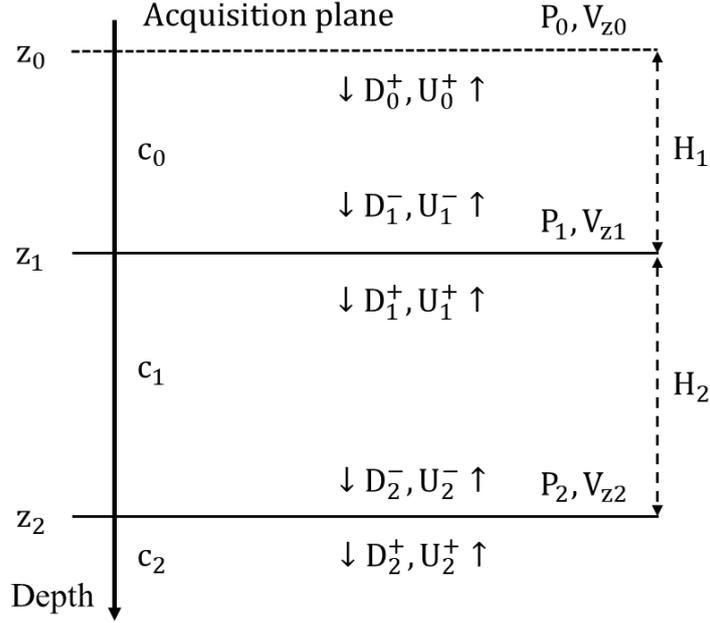

**Figure 1**: Wavefields in a 1D layered model. $c_0$, $c_1$, … are layer velocities. $H_1$, $H_2$ are the layer thicknesses. $P_i$, $v_{zi}$ are pressure and vertical component of the particle velocity at the depth $z_i$, respectively. $U_i$, $D_i$ are the up-going and down-going pressure fields respectively at depth $z_i$, where "-" indicates the wavefields on the top side of $z_i$, "+" indicates the wavefields on the bottom side of $z_i$.

To illustrate the DWI process, we assume that both the pressure waveform, P, and particle velocity, $V_z$, are recorded on the surface ($z_0$). The incident plane wave is vertical or at zero incident angle. We further assume that $c_0$, the velocity of the first layer, is known. We decompose the wavefields (P & $V_z$) into up-going, U, and down-going, D, pressure wavefields, respectively (Liu and Zheng, 2015) as follows

$$D+U = P \quad , \tag{1}$$
$$D-U = \rho c V_z \quad , \tag{2}$$

where $\rho$ and $c$ are density and wave velocity, respectively. Conversely, if we know U and D, we can compose P and $V_z$. In this section, we assume a constant density $\rho$. In the next section, the density is allowed to vary with depth.

DWI consists of four steps:

**Step 1.** At the acquisition plane depth $z_0$, we use the recorded waveforms, $P_0(t)$ and $V_{z0}(t)$, and the first layer's velocity $c_0$ (known), to calculate the up-going and down-going pressure fields at depth $z_0^+$, denoted as $U_0^+$ and $D_0^+$, respectively, using equation (1). We can view the medium below $z_0$ as a single system. In the first layer, $D_0^+$ is the incident wave, and $U_0^+$ is the reflection response of the system. Following the time-space causality of the wavefield, the earliest up-going impulse in $U_0^+$ in the first layer must be generated from the first (earliest) down-going impulse in $D_0^+$, reflected by the reflector at $z_1$. We can measure the time difference, $2\tau$, between the first impulse in $U_0^+$ and the first impulse in $D_0^+$. Hence the depth of the reflector $z_1$, or the thickness of the first layer $H_1$, can be calculated by multiplying $c_0$ with the one-way traveltime $\tau$.

**Step 2.** We then extrapolate fields $U_0^+$ and $D_0^+$ to the bottom of the first layer (depth $z_1^-$) in the frequency $\omega$-domain

$$U_1^- = U_0^+ \exp(-i\omega\tau), \qquad (3)$$

$$D_1^- = D_0^+ \exp(+i\omega\tau). \qquad (4)$$

**Step 3.** After extrapolation, the first impulse in $U_0^-$ and the first impulse in $D_0^-$ should be time-shifted to the same time as if the incidence and reflection occur right above $z_1$. We use their amplitude ratio, $R_0$, which is the reflectivity in a constant-density medium

$$R_0 = \frac{c_1 - c_0}{c_1 + c_0}, \qquad (5)$$

to determine the velocity, $c_1$, of the second layer since $c_0$ is known.

**Step 4.** Finally, we can use $c_1$, $U_1^-$, and $D_1^-$, to obtain the pressure and particle velocity fields, $P_1$ and $v_{z1}$, respectively at depth $z_1^-$, using equations (1) (2). Because the pressure and the particle velocity fields should be continuous across a boundary, we can get their values at $z_1^+$ in layer 2. At this point, we also know $c_1$, so our situation is the same as in Step 1.

The aforementioned process, using the recorded fields, $P_0$ and $V_{z0}$, and $c_0$, to obtain the other parameters of the second layer ($z_1$, $P_1$, $V_{z1}$, and $c_1$), can be recursively repeated downward layer by layer. As the inversion goes deeper, there will be fewer and fewer remaining seismic events in both the up-going and down-going data. Eventually, DWI stops when there are no up-going events in the extrapolated fields due to finite recording time of the seismic traces. We can see that DWI always converges, unconditionally.

## Simultaneous DWI for both velocity and density

In the previous 1D DWI scheme, we assume the density is constant throughout the model. For the 1D inversion on models of depth dependent density profiles, there were some relevant work by Coen in 1980s (Coen, 1981a; b; c). In Coen's work, the density and velocity are inverted separately using a dataset from oblique incident plane waves based on the Gel'fand-Levitan-Marchenko (GLM) theory (Agranovich and Marchenko, 1963; Berryman and Greene, 1980). In our study, instead of applying the GLM theory, we directly use the incident angle (θ)-dependence of the reflectivity, R(θ), to invert for both velocities and densities of a layered model. To achieve simultaneous inversion of velocities and densities, we show how to modify the steps in the previous section respectively.

Assuming the wave is incident from medium-1 ($\rho_1, c_1$) at an angle θ to medium-2 ($\rho_2, c_2$), we have the angle-dependent reflectivity

$$R(\theta) = \frac{\rho_2 c_2 \cos\theta - \rho_1 \sqrt{c_1^2 - c_2^2 \sin^2\theta}}{\rho_2 c_2 \cos\theta + \rho_1 \sqrt{c_1^2 - c_2^2 \sin^2\theta}} \quad . \tag{6}$$

If we have two plane waves of two different incident angles $\alpha$ and $\beta$ and two measured amplitude ratios $R_\alpha = R(\alpha)$ and $R_\beta = R(\beta)$, we can in principle determine $c_2$, and $\rho_2$, simultaneously.

To further improve the inversion, we can make use of data from multiple incident angles ($n \geq 2$), and minimize the objective function

$$J[R(\theta_i), r_i] = \sum_{i=1}^{n} |R(\theta_i) - r_i|^2 \quad . \tag{7}$$

In equation (7), for a plane wave at the incident angle $\theta_i$, $R(\theta_i)$ is calculated using equation (6), and $r_i$ is the measured amplitude ratio of the up-going and down-going pressure fields

Assuming the incident angle is $\theta$ and in order to invert for density, the steps of DWI need to be adjusted as follows:

In Step 1, equation (2), the relationship between the pressure and vertical component of the particle velocity, should be changed to

$$D - U = \rho c V_z / \cos\theta \quad . \tag{8}$$

In Step 2, the extrapolation of up-going and down-going pressure fields should be modified as

$$U_2^- = U_1^+ \exp(-i\omega\tau\cos\theta), \tag{9}$$
$$D_2^- = D_1^+ \exp(+i\omega\tau\cos\theta). \tag{10}$$

In Step 3, using the amplitude ratio $R(\theta_i)$ and the measured data, $r_i$, from multiple incident angles, $\theta_i$, we can obtain $c_2$ and $\rho_2$ by either solving equation (6) directly or fitting equation (7). Compared with equation (6), equation (7) uses information from multiple incident angles and can better handle errors from the data.

In Step 4, we need to use equations (1) and (8) to compose P and $V_z$ from the up-going and the down-going fields.

## Numerical Examples

In this section, we will present two synthetic examples to demonstrate the effectiveness of our proposed method: a simple layered model with six layers, and a more complex layered model with thirty-one layers. Both models are horizontally stratified. Within each layer, the velocity and density are constant. However, different layers have different properties. Both the top and bottom boundaries of the model are set up as half-space boundary conditions.

The synthetic data (pressure & particle velocity) in both examples are generated by a propagator matrix method (e.g., Eftekhar *et al.*, 2018). The plane wave is injected at a depth of 0 m and propagated downward. The receivers are placed at the same depth. Both the pressure and particle velocity wavefields are recorded at a time sampling interval of 1ms.

### Example 1

In the first example, there are six layers (Figure 2) and the velocity contrast is up to 200%. Here we use a 15 Hz Ricker wavelet as the incident plane wave for the model (Figure 2). We conduct the modeling for four plane wave sources at different incident angles: 0, 5, 10, and 15 degrees. The waveform records of the pressure and the vertical component of particle velocity are shown in Figure 3.

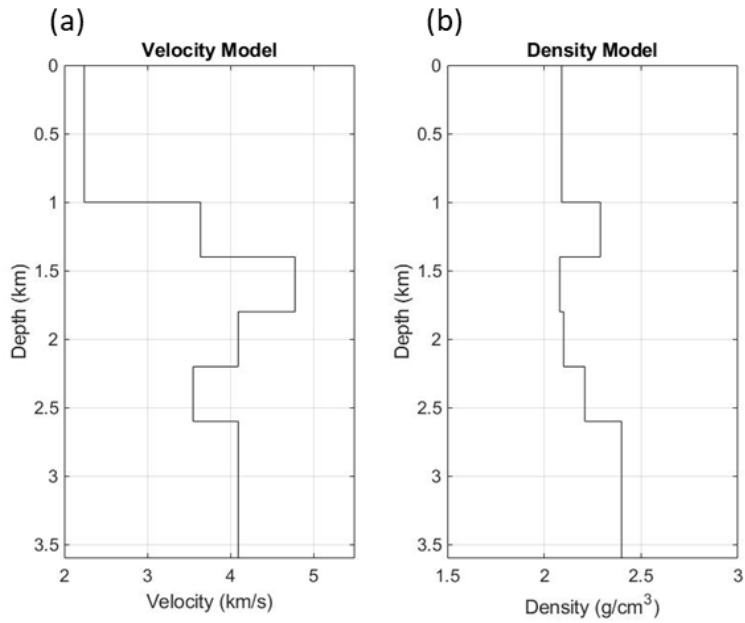

**Figure 2**: Velocity (a) and density (b) profiles of the true model.

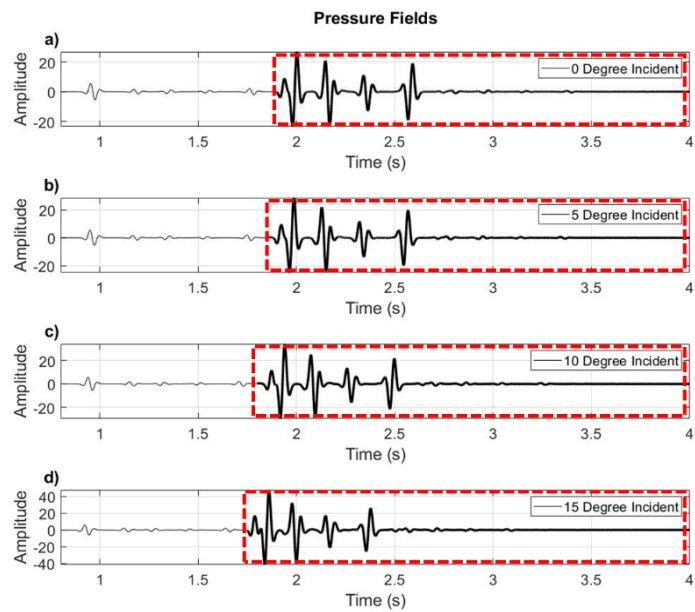

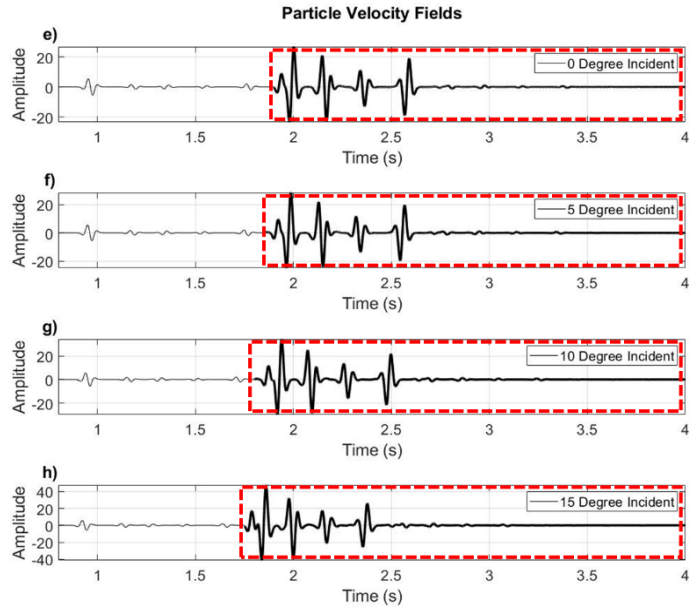

**Figure 3**: Recorded waveforms of pressure (a)-(d) and vertical component of particle velocity (e)-(h) in response to four different plane waves. The amplitudes within the red dashed boxes are amplified by 300 times.

In Figure 3, the recorded waveforms contain full information of the wavefields, including the primary reflections and multiples.

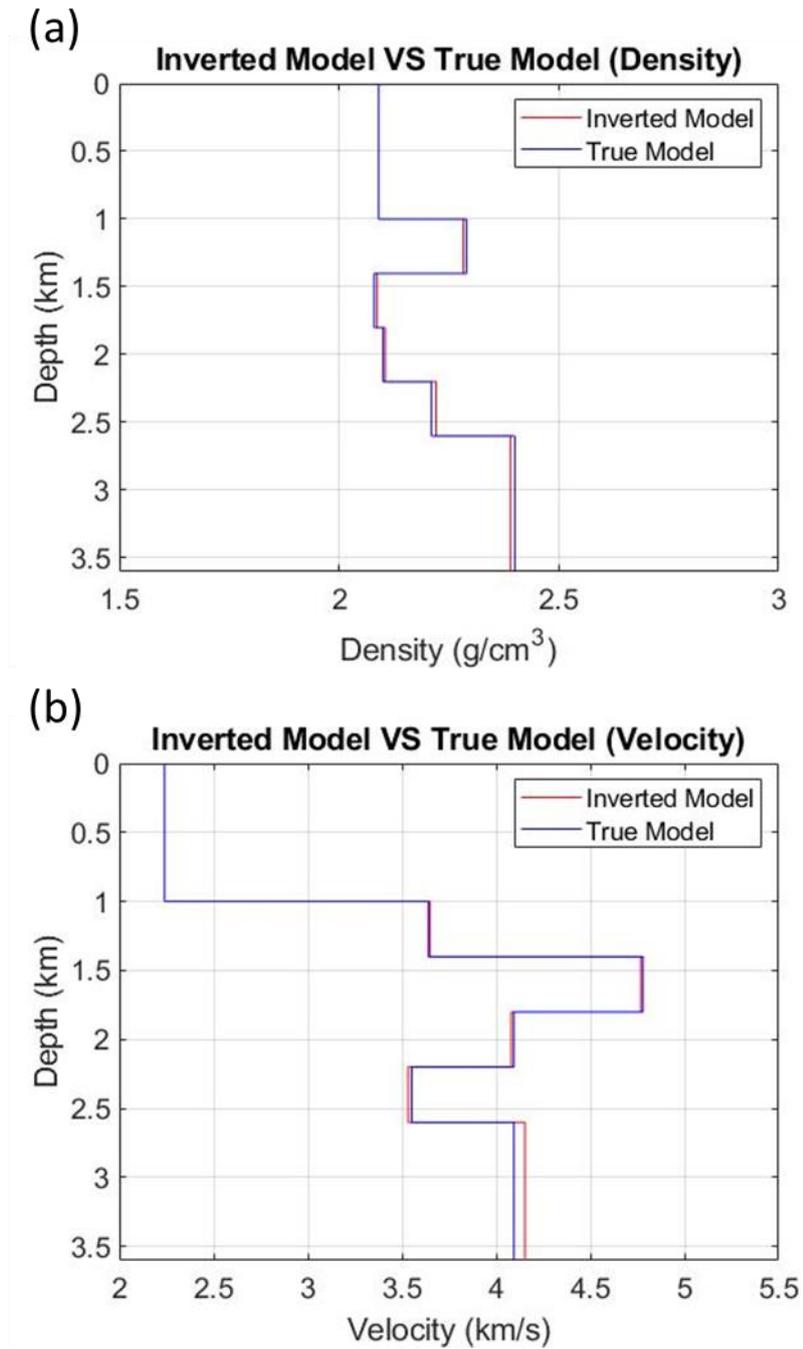

**Figure 4**: Comparisons between DWI inversion result and true model for both velocity (a) and density (b) structure.

Using the input data shown in Figure 3 and following the DWI steps in the previous section, we invert for both the velocity and density profiles, shown in Figure 4. We also calculated the misfits in velocities and densities between the DWI results

and the true models in

Table **1**. Most of them are less than 1%, except the velocity in the last layer (layer 6).

**Table 1:** The misfit of velocities and densities between the inverted result and correct model (the velocity and density information of the first layer are known).

|         | Misfit of Velocity | Misfit of Density |
|---------|--------------------|-------------------|
| Layer 2 | 0.28%              | 0.32%             |
| Layer 3 | 0.26%              | 0.29%             |
| Layer 4 | 0.34%              | 0.23%             |
| Layer 5 | 0.54%              | 0.47%             |
| Layer 6 | 1.33%              | 0.44%             |

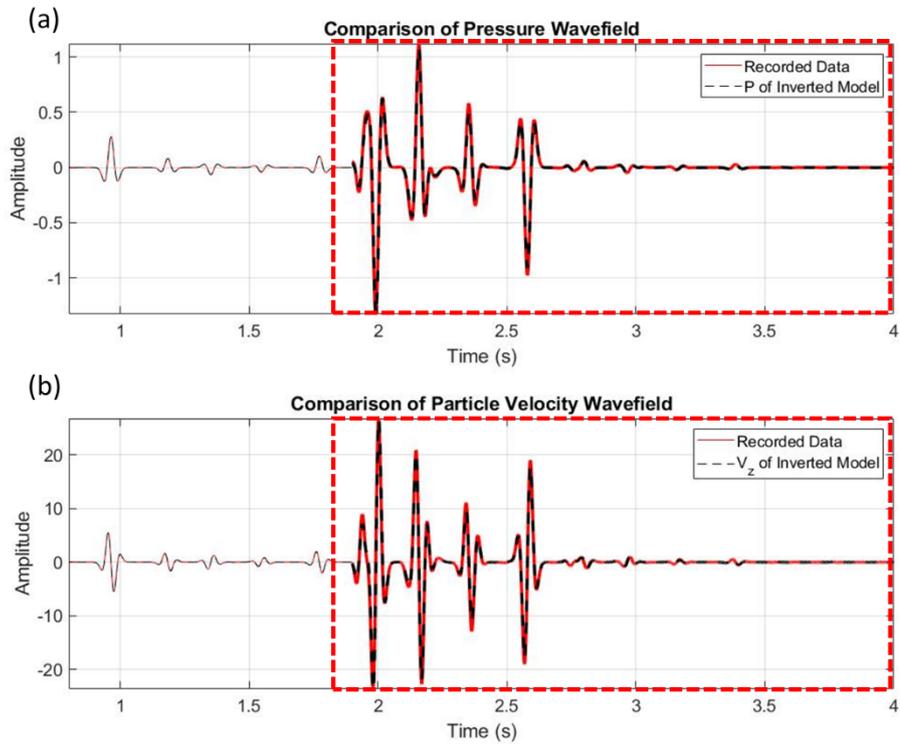

**Figure 5**: Comparisons of the recorded data (red) and synthetics (black) modeled using the DWI inverted model. (a) pressure waveforms; (b) particle velocity waveforms at the 0-degree incidence. Note the waveform amplitudes of the multiples in the red dashed box are amplified by 300 times.

To check the validity of the inverted model in the data space, we conduct a forward synthetic modeling using the DWI inverted model. The modeled waveforms fit the data very well (Figure 5). Both the primary reflections and the internal multiples can be reproduced by the DWI inverted model.

**Example 2**

In the second example, we build a model with 31 velocity and density discontinues (Figure 6). An 80 Hz Ricker wavelet is used as the incident plane wave source wavelet and we conducted modeling for four plane wave sources at angles: 0, 5, 9, and 16 degrees. The waveform recordings of the pressure and the vertical component of particle velocity are shown in Figure 7.

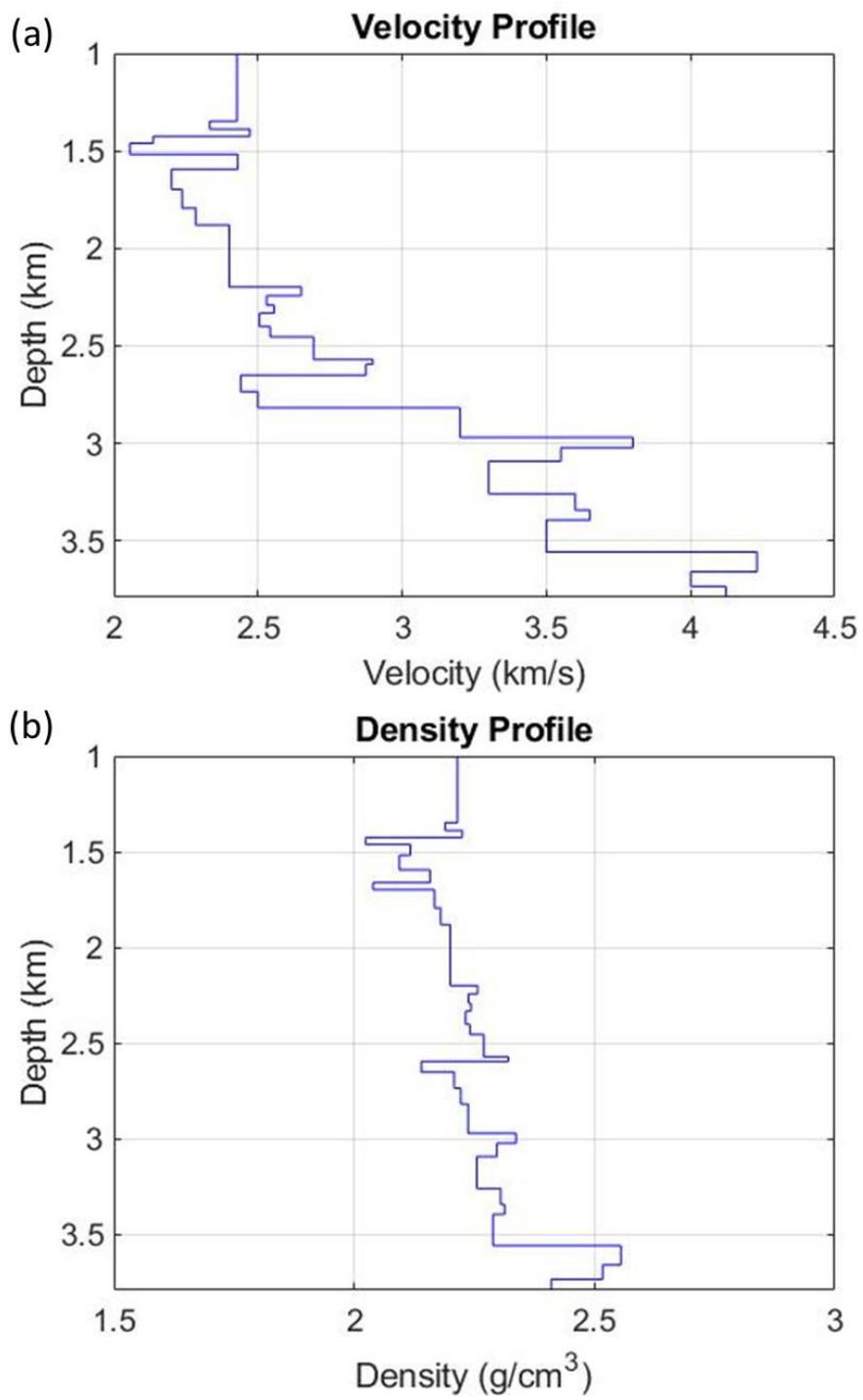

**Figure 6**: The velocity (a) and density (b) profiles of the true model for DWI.

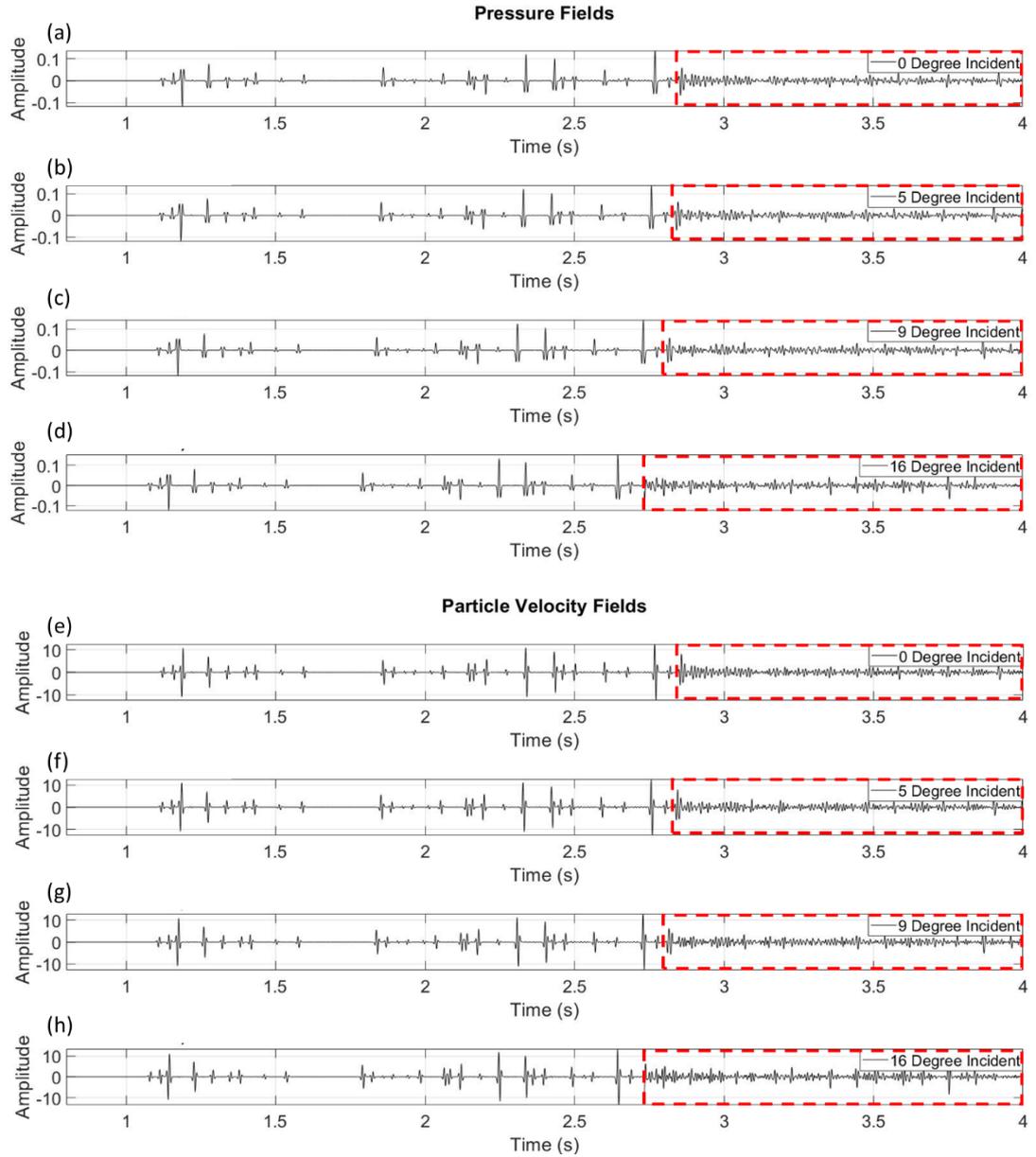

**Figure 7**: Recorded waveforms of pressure (a)-(d) and vertical component of particle velocity (e)-(h) in response to four different plane waves. The events in the red dashed boxes are amplified by 10 times.

Compared with the recorded waveforms in the first example (Figure 3), both the primary reflections and the internal multiples (Figure 7) are much more complicated. Using these data, we apply our DWI scheme and obtain the inversion results of velocity, density, and impedance models shown in Figure 8.

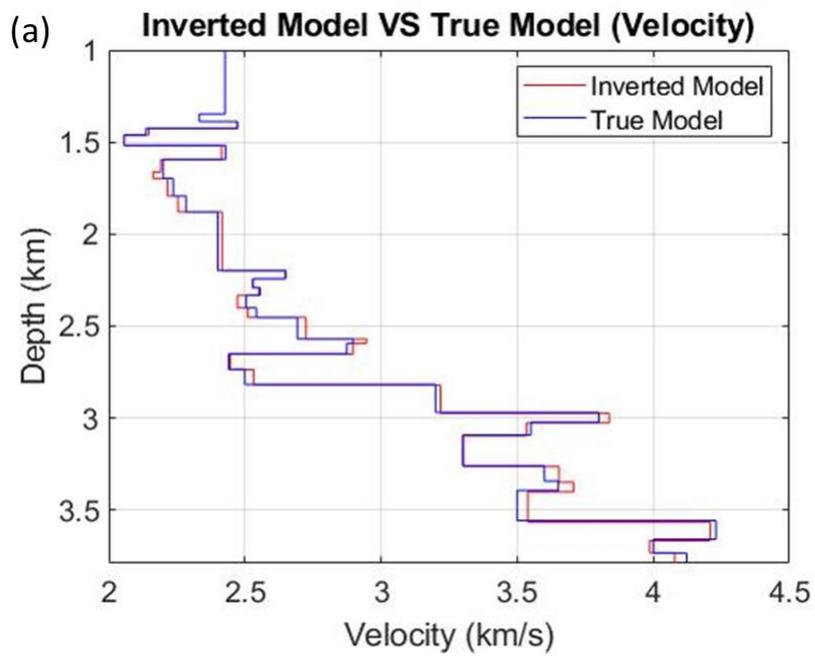

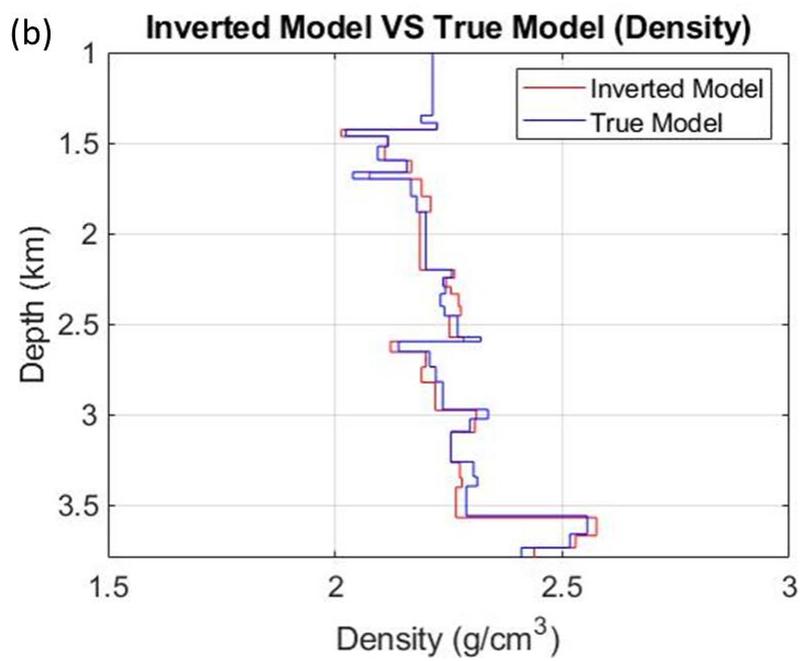

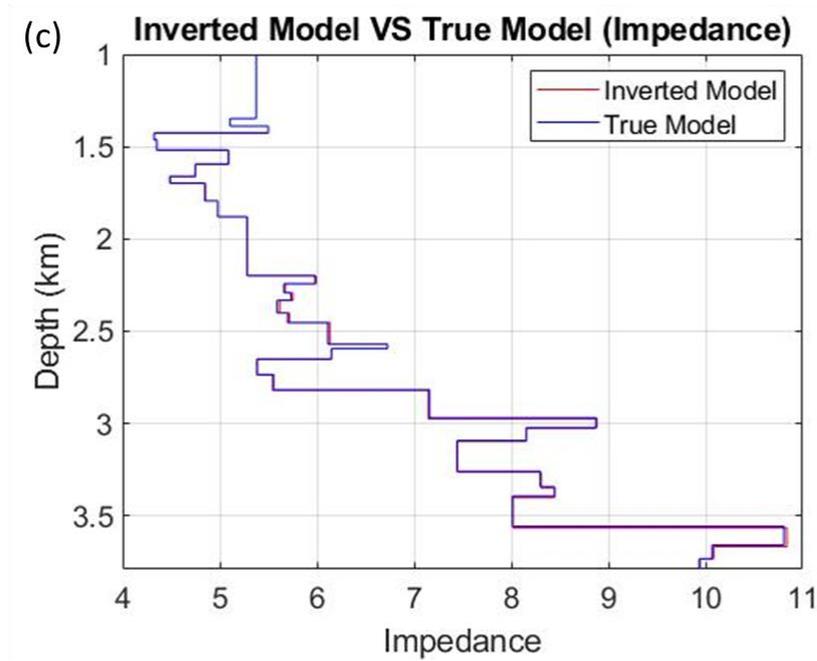

**Figure 8**: Comparisons between DWI inversion result and true model on velocity (a), density (b), and impedance (c) models.

From Figure 8 we can see the DWI scheme almost fully recovers the impedance model. For the velocity and density models, although there are some small misfits (less than 2%), the inverted models still agree well with the true model. To further examine the influences of these misfits, here we also conduct a forward synthetic modeling same as example 1 based on the inversion result, the results are shown in Figure 9. The modeled waveforms using the DWI model fit the data very well including the primary reflections and the internal multiples.

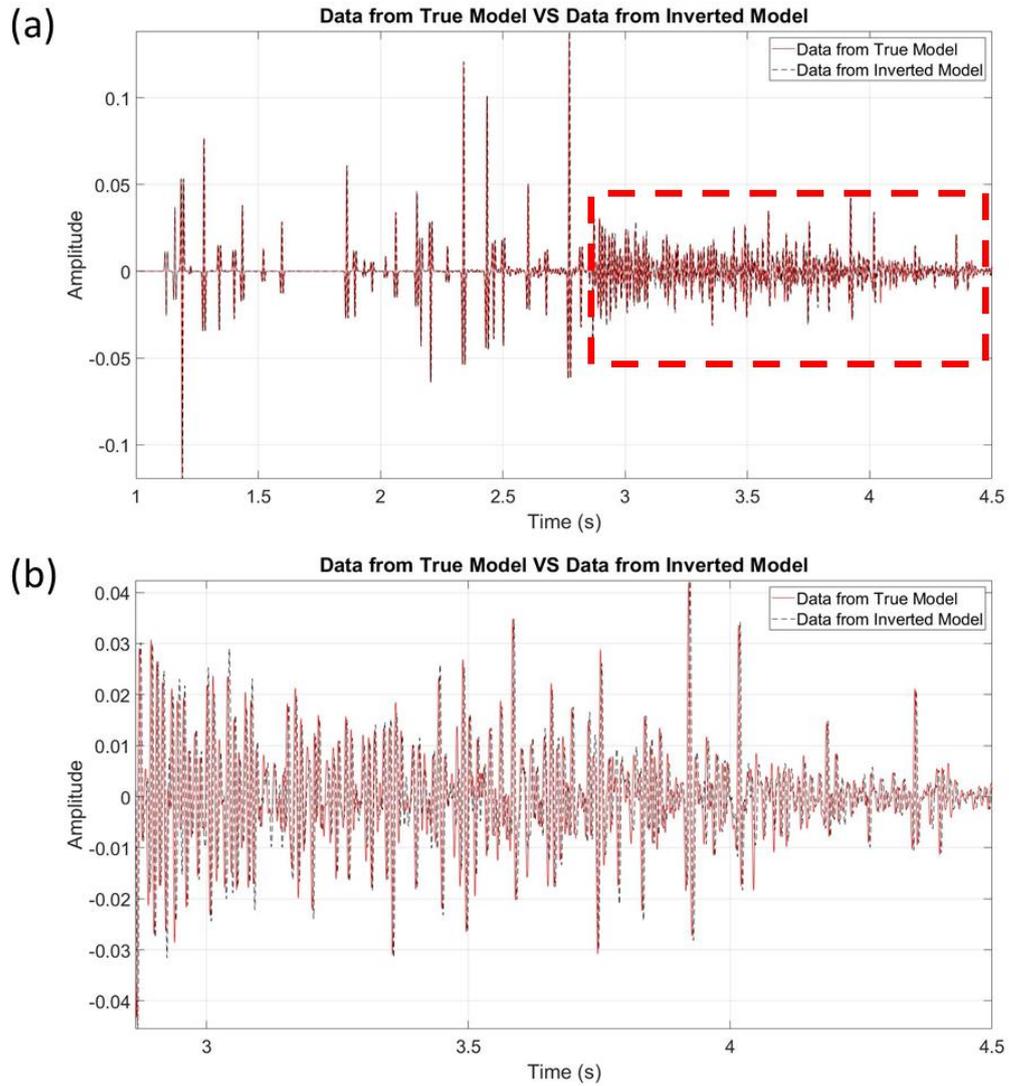

**Figure 9**: Comparisons of data (red) and synthetics (black) modeled using the DWI inverted model for pressure at the 0-degree incidence (a), the events in the red dashed box are amplified by 10 times. A zoom-in view of the events in the red dashed box in (a) is shown in (b).

## Discussion

There are significant differences between DWI and FWI. DWI does not rely on an initial global model to start the waveform inversion process. It only needs the local velocity around the receivers. For the layered model with constant velocity in each layer, the DWI scheme converts the FWI global optimization problem into many localized reflectivity inversions by invoking the causality principle. Hence it reduces the nonlinearity significantly. In DWI, large-amplitude signals from strong-contrast anomalies are desired. However, in FWI, it is not the case because if the misfit is large between the recorded data and the modeled data from the initial model, it means the intial model is far from the true model and FWI convergence can be a problem (Sirgue and Pratt, 2004).

There are also apparent differences between DWI and the 1D impedance inversion using the GLM theory (Wu and He, 2020). Most developments of the GLM theory are aiming at imaging and redatuming methods (e.g., Broggini *et al.*, 2014; Wapenaar *et al.*, 2014; van der Neut *et al.*, 2015). Wu and He (2020) tried to invert for the whole model in two-way traveltime then convert time to depth. For 1D GLM problem, the time to depth conversion can be carried out by the Liouville transform. But for 2D and 3D spatial problems, a macro-velocity model need to be used and should be obtained *a prior*. However, for DWI, the inversion is localized in a shallow to deep fashion and the inverted model is automatically obtained in the depth domain and there is no need to use a global velocity model. In future we will show 2d inversion results in which the wavefield extrapolation is much more involved using integral equations.

## Conclusion

We propose a new DWI scheme to invert for the subsurface density and velocity simultaneously, using several plane waves. Using recorded seismic data on the surface, our method inverts for the model parameters locally and recursively from shallow to deep depths. Numerical examples demonstrate the feasibility of the DWI approach to invert for both velocity and density.

# References

Agranovich, Z., and V. A. Marchenko (1963), The inverse problem of scattering theory.

Alkhalifah, T., and R.-É. Plessix (2014), A recipe for practical full-waveform inversion in anisotropic media: An analytical parameter resolution study, *Geophysics*, *79*(3), R91-R101.

Berryman, J. G., and R. R. Greene (1980), Discrete inverse methods for elastic waves in layered media, *Geophysics*, *45*(2), 213-233.

Bourgeois, A., B. F. Jiang, and P. Lailly (1989), Linearized inversion: a significant step beyond pre-stack migration, *Geophys. J. Int.*, *99*(2), 435-445.

Broggini, F., R. Snieder, and K. Wapenaar (2014), Data-driven wavefield focusing and imaging with multidimensional deconvolution: Numerical examples for reflection data with internal multiples, *Geophysics*, *79*(3), WA107-WA115.

Brossier, R., S. Operto, and J. Virieux (2009), Seismic imaging of complex onshore structures by 2D elastic frequency-domain full-waveform inversion, *Geophysics*, *74*(6), WCC105-WCC118.

Chen, G. X., R. S. Wu, and S. C. Chen (2018), Reflection multi-scale envelope inversion, *Geophysical Prospecting*, *66*(7), 1258-1271.

Claerbout, J. F. (1976), *Fundamentals of geophysical data processing*, Citeseer.

Coen, S. (1981a), Density and compressibility profiles of a layered acoustic medium from precritical incidence data, *Geophysics*, *46*(9), 1244-1246.

Coen, S. (1981b), On the elastic profiles of a layered medium from reflection data. Part I. Plane-wave sources, *The Journal of the Acoustical Society of America*, *70*(1), 172-175.

Coen, S. (1981c), On the elastic profiles of a layered medium from reflection data. Part II: Impulsive point source, *The Journal of the Acoustical Society of America*, *70*(5), 1473-1479.

Eftekhar, R., H. Hu, and Y. J. G. J. I. Zheng (2018), Convergence acceleration in scattering series and seismic waveform inversion using nonlinear Shanks transformation, *Geophysical Journal International*, *214*(3), 1732-1743.

Gauthier, O., J. Virieux, and A. Tarantola (1986), Two-dimensional nonlinear inversion of seismic wave-forms - numerical results, *Geophysics*, *51*(7), 1387-1403.

Goupillaud, P. L. (1961), An approach to inverse filtering of near-surface layer effects from seismic records, *Geophysics*, *26*(6), 754-760.

Jeong, W., H.-Y. Lee, and D.-J. Min (2012), Full waveform inversion strategy for density in the frequency domain, *Geophysical Journal International*, *188*(3),

1221-1242.

Kim, Y., C. Shin, H. Calandra, and D.-J. Min (2013), An algorithm for 3D acoustic time-Laplace-Fourier-domain hybrid full waveform inversion, *Geophysics*, *78*(4), R151-R166.

Lailly, P.(1983), The Seismic Inverse Problem as a Sequence of Before-Stack Migrations, *Conference on Inverse Scattering: Theory and Applications, SIAM*,

Liu, Y., B. He, H. Lu, Z. Zhang, X. Xie, and Y. Zheng (2018), Full intensity waveform inversion, *Geophysics*, *86*(6), R649-R658.

Liu, Y., B. He, Z. Zhang, Y. Zheng, and P. Li (2020), Reflection intensity waveform inversion, *Geophysics*, https://doi.org/10.1190/geo2019-0590.1191.

Liu, Z., and Y. Zheng (2015), Direct waveform inversion, in *SEG Technical Program Expanded Abstracts 2015*, edited,vol. pp. 1268-1273,SEG Technical Program Expanded Abstracts,Series Vol. Series editor Society of Exploration Geophysicists.

Liu, Z., and Y. Zheng (2017), Applications of the direct-waveform inversion on 2D models, in *SEG Technical Program Expanded Abstracts 2017*, edited,vol. pp. 1687-1691Vol. Series editor Society of Exploration Geophysicists.

Luo, J., and R. S. Wu (2015), Seismic envelope inversion: reduction of local minima and noise resistance, *Geophysical Prospecting*, *63*(3), 597-614.

Mora, P. (1987), Nonlinear two-dimensional elastic inversion of multioffset seismic data, *Geophysics*, *52*(9), 1211-1228.

Richardson, A. (2018), Seismic full-waveform inversion using deep learning tools and techniques.

Sears, T. J., S. Singh, and P. Barton (2008), Elastic full waveform inversion of multi-component OBC seismic data, *Geophysical Prospecting*, *56*(6), 843-862.

Shin, C., and Y. H. Cha (2008), Waveform inversion in the Laplace domain, *Geophysical Journal International*, *173*(3), 922-931.

Shin, C., and W. Ha (2008), A comparison between the behavior of objective functions for waveform inversion in the frequency and Laplace domains, *Geophysics*, *73*(5), VE119-VE133.

Sirgue, L., and R. G. Pratt (2004), Efficient waveform inversion and imaging: A strategy for selecting temporal frequencies, *Geophysics*, *69*(1), 231-248.

Tarantola, A. (1984), Inversion of seismic-reflection data in the acoustic approximation, *Geophysics*, *49*(8), 1259-1266.

Tarantola, A. (1986), A strategy for nonlinear elastic inversion of seismic-reflection data, *Geophysics*, *51*(10), 1893-1903.

Tarantola, A. (2005), *Inverse problem theory and methods for model parameter estimation*, SIAM.

van der Neut, J., J. Thorbecke, K. Wapenaar, and E. Slob (2015), Inversion of the multidimensional Marchenko equation, *77th EAGE Conference and Exhibition 2015*, **2015**, 1-5,

Virieux, J., and S. Operto (2009), An overview of full-waveform inversion in exploration geophysics, *Geophysics*, *74*(6), WCC1-WCC26.

Wapenaar, K., J. Thorbecke, J. Van Der Neut, F. Broggini, E. Slob, and R. J. G. Snieder (2014), Marchenko imaging, *Geophysics*, *79*(3), WA39-WA57.

Warner, M., A. Ratcliffe, T. Nangoo, J. Morgan, A. Umpleby, N. Shah, V. Vinje, I. Štekl, L. Guasch, and C. Win (2013), Anisotropic 3D full-waveform inversion, *Geophysics*, *78*(2), R59-R80.

Wu, R.-S., J. Luo, and B. Wu (2014), Seismic envelope inversion and modulation signal model, *Geophysics*, *79*(3), WA13-WA24.

Wu, R.-S., and Y. Zheng (2014), Non-linear partial derivative and its De Wolf approximation for non-linear seismic inversion, *Geophysical Journal International*, *196*(3), 1827-1843.

Wu, R., and H. He (2020), Inverse Scattering for Schrödinger Impedance Equation and Simultaneous ρ-v Inversion in Layered Media, *82nd EAGE Annual Conference & Exhibition*, **2020**, 1-5,

Zheng, Y., and Z. Liu (2020), Concepts in the Direct Waveform Inversion (DWI) Using Explicit Time-Space Causality, *COMMUNICATIONS IN COMPUTATIONAL PHYSICS 28*(1), 342-355.